\documentclass[aps,preprint,showpacs,amsmath,amssymb]{revtex4}
\usepackage{graphicx}
\usepackage{amssymb}
\usepackage{bm}

\begin{document}
\setcounter{page}{1}

\title{Spontaneous emission from a two-level atom in anisotropic one-band photonic crystals: a fractional calculus approach}

\author{Jing-Nuo \surname{Wu$^{1}$}}
\author{Chih-Hsien \surname{Huang$^{1}$}}
\author{Szu-Cheng \surname{Cheng$^{2}$}}
\email {sccheng@faculty.pccu.edu.tw}
\thanks{FAX: +886-2-28610577}
\author{Wen-Feng \surname{Hsieh$^{1,3}$}}
\email {wfhsieh@mail.nctu.edu.tw}
\thanks{FAX: +886-3-5716631}
\affiliation{$^{1}$Department of Photonics and Institute of Electro-Optical Engineering, National Chiao Tung University, Hsinchu, Taiwan, R. O. C.}
\affiliation{$^{2}$Department of Physics, Chinese Culture University, Taipei, Taiwan, R. O. C.}
\affiliation{$^{3}$Department of Electro-Optical Engineering, National Cheng Kung University, Tainan, Taiwan, R. O. C.}
\date[]{Received \today }

\begin{abstract}
Spontaneous emission (SE) from a two-level atom in a photonic crystal (PC) with anisotropic one-band model is investigated using the fractional calculus. Analytically solving the kinetic equation in terms of the fractional exponential function, the dynamical discrepancy of SE between the anisotropic and isotropic systems is discussed on the basis of different photon density of states (DOS) and the existence of incoherent diffusion field that becomes even more clearly as the atomic transition frequency lies close to the band edge. With the same atom-field coupling strength and detuning in the forbidden gap, the photon-atom bound states in the isotropic system turn into the unbound ones in the anisotropic system that is consistent with the experimental observation in $Phys.$ $Rev.$ $Lett.$ \textbf{96}, 243902 (2006). Dynamics along different wavevectors with various curvatures of dispersion is also addressed with the changes of the photon DOS and the appearance of the diffusion fields. 
\end{abstract}

\pacs{05.40.-a , 42.50.-p, 32.80.-t}
\maketitle

\section{INTRODUCTION}
Photonic crystals (PCs), a new class of optical materials with periodic dielectric structure, provide a way to control spontaneous emission (SE) through redistributing the photon density of states (DOS) near photonic band gap (PBG). This control offers the key technology of manipulating light, such as in light emitting devices \cite{LChen04}, quantum information processing \cite{NVats01}, or solar cells \cite{SNishimura03}. The special photon density of states near band edge changes the optical behavior of an atom in a photonic crystal including the appearance of photon-atom bound states \cite{SJohn90, SJohn91, SJohn97, SBay97, SCCheng09}, spectral splitting \cite{SJohn94}, enhanced quantum interference effects \cite{SYZhu97}, coherent control of SE \cite{TQuang97}, non-Markovian effects \cite{SJohn90, SJohn91, SJohn97, AGKofman94}, etc. Among these studies, the photon reservoir of a photonic crystal could be well described by one band edge frequency $\omega_{c}$ at a certain edge wavevector and a dispersion relation which leads to the special photon DOS. The dispersion relation near the band edge is assumed to be isotropic in early studies \cite{SJohn90, SJohn91, SJohn97, SBay97, SJohn94, AGKofman94}. In a real three-dimensional photonic crystal with an allowed point-group symmetry, the photonic band structure is highly anisotropic, namely, the equal frequency surface near the band edge is no longer spherical. A vector form of photon dispersion relation is required to describe a more realistic picture of the band edge behavior. In this vector form of dispersion relation, the photon DOS is proportional to $\sqrt{\omega-\omega_{c}}$ while that of an isotropic dispersion relation is proportional to $1/\sqrt{\omega-\omega_{c}}$, where $\omega$ stands for the eigenmode frequency and $\omega_{c}$ for the band edge frequency \cite{SJohn95}. This discrepancy of DOS will cause dynamical difference of spontaneous emission between the anisotropic and isotropic systems. The most prominent difference is the existence of diffusion field \cite{SYZhu00, YYang03A, YYang03, YYang06}. It was predicted theoretically by Yang et al. \cite{SYZhu00, YYang03A, YYang03, YYang06} that the bare atomic transition frequency lying in the region near band edge will be shifted into the forbidden gap by the interaction with the radiation modes where an atom-photon bound state is generated. That is, the emitter with frequency lying in the forbidden gap will not yield SE. Recently, Barth et al. observed experimentally \cite{MBarth06} that the anisotropic properties of a PC could be detected by employing single emitters such as quantum dots (QDs). When an emitter is placed inside a PC with the anisotropic band structure, the additional anisotropy will imprint on the spontaneous emission of the system if the emission frequency lies in the forbidden gap. In this observation, the CdSe/ZnS quantum dots embedded inside artificial colloidal opals with direction-dependent band structure gave fluorescence image (SE) with an extra anisotropy but did not emit light if embedded inside a PC with weak anisotropy of band structure for the emission frequency of the QDs lying in the forbidden gap. There exists inconsistency between the theoretical prediction and experimental observation because SE appears in the anisotropic PC system even for the emitter's frequency lying in the forbidden gap. These inconsistency arouses our attention for the physical properties of the anisotropic PC system. We found that the presence of SE from the QDs in the anisotropic PC results from the dynamical difference of SE from the isotropic (or weak anisotropic) PC.

In this paper, we study the dynamics of SE from a two-level atom embedded in a PC with anisotropic one-band model (Fig. 1). Fractional calculus is applied to solve the non-Markovian dynamics of the anisotropic optical system with threshold-like DOS. We found that the dynamical discrepancy of SE between the anisotropic and isotropic systems is the consequence of the different DOS in these two systems and the existence of the diffusion field in the anisotropic system. With the same atom-field coupling strength and detuning in the forbidden gap, the photon-atom bound states near the band edge of the isotropic reservoir turn into the decaying states in anisotropic system. This alteration leads to the presence of SE in the anisotropic PC system which is consistent with the experimental observation by Barth et al. \cite{MBarth06}. The new topic of how the curvature of the anisotropic dispersion relation affects the dynamical behavior of the system has also been investigated. The dynamical difference between the systems with various curvatures of dispersion is observed and discussed on the basis of the curvature-dependent DOS and coupling strength. 

The paper is organized as follows. In Sec. II, the basic theory of the anisotropic system is given. In Sec. III, we solve the kinetic equation of the anisotropic system through fractional calculus and express the analytical solution in terms of the fractional exponential function. The dynamical behavior of the anisotropic system is compared with that of the isotropic system. And the influence of curvature of the anisotropic dispersion relation on the dynamical behavior is also discussed here. Finally, we summarize our results in Sec. IV.

\section{Basic theory of a two-level atom in an anisotropic one-band photonic crystal}
When the system of a two-level atom coupled to the field reservoir in a photonic crystal with anisotropic one-band model is considered, the Hamiltonian for this atom-field interacting system can be written as     
\begin{equation}	H=\hbar\omega_{21}\sigma_{22}+\sum_{\vec{k}}{\hbar\omega_{\vec{k}}a_{\vec{k}}^{+}a_{\vec{k}}+i\hbar\sum_{\vec{k}}{g_{\vec{k}}(a_{\vec{k}}^{+}\sigma_{12}-\sigma_{21}a_{\vec{k}})}},
\end{equation}
where $\sigma_{ij}=|{i}\rangle\langle {j}|$ (\textit{i},\textit{j}=1,2) are the atomic operators for a two-level atom with excited state $|{2}\rangle$, ground state $|1\rangle$, and resonant transition frequency $\omega_{21}$; $a_{\vec{k}}$ and $a_{\vec{k}}^{+}$ are the annihilation and creation operators of the radiation field; $\omega_{\vec{k}}$ is the radiation frequency of mode $\vec{k}$ in the reservoir, and the atom-field coupling constant $g_{\vec{k}}=\frac{\omega_{21}d_{21}}{\hbar}[{\frac{\hbar}{2\epsilon_{0}\omega_{\vec{k}}V}]}^{\frac{1}{2}}\hat{e}_{\vec{k}}\cdot\hat{u}_d$ is assumed to be independent of atomic position with the fixed atomic dipole moment $\vec{d}_{21}=d_{21}\hat{u}_d$. Here $V$ is the sample volume, $\hat{e}_{\vec{k}}$ is the polarization unit vector of the reservoir mode $\vec{k}$, and $\epsilon_{0}$ is the Coulomb constant.

In the single photon sector, the wave function of the system has the form
\begin{equation}
	|\psi(t)\rangle=B(t)e^{-i\omega_{21}t}|2,\{0\}\rangle+\sum_{\vec{k}}{C_{\vec{k}}(t)e^{-i\omega_{\vec{k}}t}|1,\{1_{\vec{k}}\}\rangle}
\end{equation}
with initial condition $B(0)=1$ and $C_{\vec{k}}(0)=0$.  Here $B(t)$ labels the probability amplitude for the atom in its excited state $|2\rangle$ with an electromagnetic vacuum state and $C_{\vec{k}}(t)$ for the atom in its ground state $|1\rangle$ with a single photon in mode $\vec{k}$ with frequency $\omega_{\vec{k}}$. We got the equations of motion for the amplitudes by projecting the time-dependent Schr$\ddot{o}$dinger equation on the one-photon sector of the Hilbert space as
\begin{equation}
\frac{d}{dt}B(t)=-\sum_{\vec{k}}{g_{\vec{k}}C_{\vec{k}}(t)e^{-i\Omega_{\vec{k}}t}}
\end{equation}
\begin{equation}
\frac{d}{dt}C_{\vec{k}}(t)=g_{\vec{k}}B(t)e^{i\Omega_{\vec{k}}t}
\end{equation}
with detuning frequency $\Omega_{\vec{k}}=\omega_{\vec{k}}-\omega_{21}$. By substituting the time integration of equation (4) into equation (3), we have the time evolving equation of the excited-state probability amplitude
\begin{equation}
\frac{d}{dt}B(t)=-\int_{0}^{t}{G(t-\tau)B(\tau)d\tau}
\end{equation}
with the memory kernel 
$G(t-\tau)=\sum_{\vec{k}}{g_{\vec{k}}^{2}e^{-i\Omega_{\vec{k}}(t-\tau)}}$. This memory kernel is related to the dispersion relation of photon reservoir. For the anisotropic photonic band gap (PBG) reservoir, the dispersion relation has a vector form and could be expressed by the effective-mass approximation as \cite{SJohn87} 
\begin{equation} 
\omega_{\vec{k}}\approx\omega_{c}+A\left(\vec{k}-\vec{k}_{c}\right)^{2}
\end{equation} 
where $A\cong f\omega_{c}/k^{2}_{c}$ signifies different curvatures in different directions with scaling factor $f$ whose value depends on the nature of the dispersion relation near the band edge $\omega_{c}$. The related memory kernel of the anisotropic system could be further expressed as
\begin{equation}
G(t-\tau)=\frac{\omega^{2}_{21}d^{2}_{21}}{12\pi^{2}\epsilon_{0}\hbar\omega_{c}}\int\rho(\omega)e^{-i(\omega-\omega_{21})(t-\tau)}d\omega
\end{equation}
with the curvature-dependent density of states \cite{SJohn90, SJohn91}
\begin{equation}
\rho(\omega)=\sqrt{\frac{\omega-\omega_{c}}{A^{3}}}\Theta(\omega-\omega_{c}),
\end{equation}
where $\Theta(\omega-\omega_{c})$ is the Heaviside step function. This density of states has threshold-like behavior near band edge $\omega_{c}$ which results in the different dynamical behavior of this atom-field interacting system from that of free space. By applying the complex Fresnel integral $\int^{\infty}_{0}x^{p-1}e^{-ax}dx=\Gamma(p)/a^{p}$ \cite{ISGradshteyn80}, we could integrate this memory kernel as 
\begin{equation}
G(t-\tau)=\frac{\beta^{1/2}/f^{3/2}}{\sqrt{\pi}(t-\tau)^{3/2}}e^{-i[3\pi/4-\Delta_{c}(t-\tau)]}
\end{equation}
with the detuning frequency $\Delta_{c}=\omega_{21}-\omega_{c}$ of the atomic resonance frequency $\omega_{21}$ from the band edge $\omega_{c}$, coupling constant $\beta^{1/2}=(\omega^{2}_{21}d^{2}_{21}\omega^{1/2}_{c})/(24\pi\epsilon_{0}\hbar c^{3})$, and $t>\tau$ in the long time limit \cite{NVats98}. Substituting this memory kernel into the time evolving equation (5) and making the transformation $B(t)=e^{i\Delta_{c}t}D(t)$, we have the kinetic equation of this anisotropic system as
\begin{equation} 
\frac{d}{dt}D(t)+i\Delta_{c}D(t)=\frac{\beta^{1/2}e^{i\pi/4}}{\sqrt{\pi}f^{3/2}}\int^{t}_{0}\frac{D(\tau)}{(t-\tau)^{3/2}}d\tau.
\end{equation}

\section{Dynamics of spontaneous emission}
In this section, we will use the mathematical method of fractional calculus to solve the kinetic equation of the anisotropic system and discuss the dynamics of SE on the basis of the obtained analytical solution.

When the right-hand-side term of the kinetic equation (10) is considered, we could express it as a Riemann-Liouville fractional differentiation operator \cite{KBOldham74, SGSamko93, KSMiller93} with order $\nu=1/2$. That is,
\begin{equation}
\int^{t}_{0}\frac{D(\tau)}{(t-\tau)^{3/2}}d\tau=\Gamma(-1/2)\frac{d^{1/2}}{dt^{1/2}}D(t)
\end{equation} 
with Gamma function $\Gamma(x)$. The kinetic equation thus has a fractional differential form as 
\begin{equation}
\frac{d}{dt}D(t)+i\Delta_{c}D(t)+\frac{2\beta^{1/2}e^{i\pi/4}}{f^{3/2}}\frac{d^{1/2}}{dt^{1/2}}D(t)=0.
\end{equation}
In order to solve this fractional kinetic equation, we manipulated the fractional operators including the integral operator ($d^{-1}/dt^{-1}$) and the fractional differentiation operator $d^{1/2}/dt^{1/2}$. Mathematically, the adopted manipulation is not unique provided that one could justify the function arrived at is the solution of the original fractional differential equation. The first step of our manipulation yielded
\begin{equation}
D(t)-D(0)+i\Delta_{c}\frac{d^{-1}}{dt^{-1}}D(t)+\frac{2\beta^{1/2}e^{i\pi/4}}{f^{3/2}}\frac{d^{-1/2}}{dt^{-1/2}}D(t)=0.
\end{equation}
The second step gave 
\begin{equation}
\frac{d^{1/2}}{dt^{1/2}}D(t)+i\Delta_{c}\frac{d^{-1/2}}{dt^{-1/2}}D(t)+\frac{2\beta^{1/2}e^{i\pi/4}}{f^{3/2}}D(t)=\frac{t^{-1/2}}{\sqrt{\pi}}.
\end{equation} 
Here the initial condition $D(0)=B(0)=1$ has been applied. The probability amplitude $D(t)$ could be solved by performing the Laplace transform of the fractional derivative including \cite{KSMiller93}
\begin{equation}
\textit{L}\left[\frac{d^{1/2}}{dt^{1/2}}D(t)\right] =s^{1/2}\tilde{D}(s)-\left[\frac{d^{-1/2}}{dt^{-1/2}}D(0)\right]_{t=0}, 
\end{equation}
\begin{equation}
\textit{L}\left[\frac{d^{-1/2}}{dt^{-1/2}}D(t)\right]=s^{1/2}\tilde{D}(s), 
\end{equation}
and 
\begin{equation}
\textit{L}[t^{-1/2}]=\frac{\Gamma(1/2)}{\sqrt{s}},
\end{equation}
where $\tilde{D}(s)$ is the Laplace transform of $D(t)$. These procedures gave the Laplace transform of $D(t)$ as
\begin{equation}
\tilde{D}(s)=\frac{1}{s+i\Delta_{c}+2\beta^{1/2}e^{i\pi/4}s^{1/2}/f^{3/2}}.
\end{equation}
In order to express this equation as a sum of partial fractions, we need to find the roots of the indicial equation $Y^{2}+2\beta^{1/2}e^{i\pi/4}Y/f^{3/2}+i\Delta_{c}=0$, where we have converted the variable $s^{1/2}$ into $Y$. There are two kinds of roots for this indicial equation: one with different roots $Y_{1}\neq Y_{2}$ and the other with degenerate root $Y_{1}=Y_{2}$. For the case of different roots, $\tilde{D}(s)$ could be expressed as 
\begin{equation}
\tilde{D}(s)=\left(\frac{1}{(\sqrt{s}-Y_{1})}-\frac{1}{(\sqrt{s}-Y_{2})}\right)\frac{1}{(Y_{1}-Y_{2})}
\end{equation}
with
\begin{equation}
Y_{1}=e^{i\pi/4}\left(-\frac{\beta^{1/2}}{f^{3/2}}+\sqrt{\frac{\beta}{f^{3}}-\Delta_{c}}\right)
\end{equation}
and
\begin{equation}
Y_{2}=e^{i\pi/4}\left(-\frac{\beta^{1/2}}{f^{3/2}}-\sqrt{\frac{\beta}{f^{3}}-\Delta_{c}}\right).
\end{equation}
For the degenerate case, we have $\frac{\beta}{f^{3}}=\Delta_{c}$. The indicial equation becomes
\begin{equation}
Y^{2}+2\frac{\beta^{1/2}e^{i\pi/4}}{f^{3/2}}Y+\frac{\beta e^{i\pi/2}}{f^{3}}=0
\end{equation}
or
\begin{equation}
\left(Y+\frac{\beta^{1/2}e^{i\pi/4}}{f^{3/2}}\right)^{2}=0.
\end{equation}
The partial fractions of $\tilde{D}(s)$ can thus be written as
\begin{equation}
\tilde{D}(s)=\frac{1}{\left(\sqrt{s}+\frac{\beta^{1/2}e^{i\pi/4}}{f^{3/2}}\right)^{2}}.
\end{equation}
The dynamical solution of the probability amplitude $D(t)$ could be obtained by applying the inverse Laplace transform on the partial-fractional forms of $\tilde{D}(s)$ for the two cases of different roots and degenerate root. The applied formula of the inverse Laplace transform include 
\begin{equation}
\textit{L}^{-1}\left[\frac{1}{(\sqrt{s}-a)}\right]=E_{t}\left(-\frac{1}{2},a^{2}\right)+aE_{t}\left(0,a^{2}\right) 
\end{equation}
and 
\begin{equation}
\textit{L}^{-1}\left[\frac{1}{(\sqrt{s}-a)^{2}}\right]=2atE_{t}\left(-\frac{1}{2},a^{2}\right)+\left(1+2a^{2}t\right)E_{t}\left(0,a^{2}\right)+aE_{t}\left(\frac{1}{2},a^{2}\right) 
\end{equation}
with $E_{t}(\alpha,a)=t^{\alpha}\sum^{\infty}_{n=0}\frac{(at)^{n}}{\Gamma(\alpha+n+1)}=\frac{d^{-\alpha}}{dt^{-\alpha}}e^{at}$ being the two-parameter fractional exponential function  of variable $t$, order $\alpha$, and constant $a$ \cite{KSMiller93}. The analytical solution for the fractional kinetic equation (Eq. (12)) of the anisotropic photonic crystal system is thus obtained.
For $\beta/f^{3}\neq\Delta_{c}$,
\begin{equation}
D(t)=\frac{1}{2e^{i\pi/4}\sqrt{\beta/f^{3}-\Delta_{c}}}\times\left[Y^{2}_{1}E_{t}(1/2,Y^{2}_{1})-Y^{2}_{2}E_{t}(1/2,Y^{2}_{2})+Y_{1}e^{Y^{2}_{1}t}-Y_{2}e^{Y^{2}_{2}t}\right].  
\end{equation}
For $\beta/f^{3}=\Delta_{c}$, 
\begin{eqnarray}
D(t)=-2\frac{\beta^{3/2}e^{i3\pi/4}}{f^{9/2}}tE_{t}(\frac{1}{2},i\beta/f^{3})-\frac{\beta^{1/2}e^{i\pi/4}}{f^{3/2}}E_{t}(\frac{1}{2},i\beta/f^{3}) \nonumber\\ 
+(1+2it\beta/f^{3})e^{i\beta t/f^{3}}-2\frac{\beta^{1/2}e^{i\pi/4}}{f^{3/2}}t^{1/2}/\sqrt{\pi}.
\end{eqnarray}
Here we have applied the relation of the fractional exponential function for special values $E_{t}(-1/2,a)=aE_{t}(1/2,a)+t^{-1/2}/\sqrt{\pi}$ and $E_{t}(0,a)=e^{at}$.
 
This analytical solution, which has so far not been obtained, determines the dynamical behavior of the atomic excitation $B(t)$ and the amplitude of the radiation field which could be obtained via $B(t)$ in a standard way \cite{MOScully97, YYang03}. Different kinds of the indicial roots $Y_{1}$ and $Y_{2}$ gave different dynamical behavior of the system which depend on the lying regions of the atomic transition frequency ( see Eq. (20) and (21)). As the atomic frequency lies in the region $\omega_{21}<\omega_{c}+\beta/f^{3}$ ($\Delta_{c}<\beta/f^{3}$), the square of these roots is pure imaginary and the dynamical solution has some non-decaying terms. Near band edge ($\omega_{21}\cong\omega_{c}$, $Y^{2}_{1}\cong Y^{2}_{2}$), these non-decaying terms interfere each other severely which leads to the decaying solution. These non-decaying terms oscillate individually with time and form atom-photon bound states as the atomic frequency lies deeply in the forbidden gap ($\omega_{21}<<\omega_{c}$). On the other hand, when the atomic frequency is in the higher-energy region $\omega_{21}>\omega_{c}+\beta/f^{3}$ (region of allowed band), the square of these roots is complex and the solution decays with time quickly. These two regions of atomic transition frequency given by the roots of the indicial equation are the discussing basis of the previous studies \cite{SYZhu00, YYang03A, YYang03, YYang06}. It was predicted in the previous studies that the bare atomic transition frequency lying in the region $\omega_{21}<\omega_{c}+\beta^{1/2}\omega^{1/2}_{c}$ will be shifted into the forbidden gap by the interaction with the radiation modes where a photon-atom bound state is generated. That is, there will not exist spontaneous emission in the anisotropic PC system if the atomic transition frequency lies in the region near band edge ($\omega_{21}\cong\omega_{c}$). This result is inconsistent with the experimental observation in $Phys.$ $Rev.$ $Lett.$ \textbf{96}, 243902 (2006) where SE appeared with an extra angular anisotropy in the anisotropic PC system as the emission frequency of the embedded QDs lies in the forbidden gap.

Instead of analyzing the integration contours for the probability amplitudes in previous solving procedures, we plot the dynamical behavior of this anisotropic system directly from the analytical solution. Based on the excited-state probability amplitude $P(t)=\left|B(t)\right|^{2}=\left|D(t)\right|^{2}$, the dynamical behavior of the anisotropic system is shown in Fig. 2. It could be observed from Fig. 2 that typical characteristic of non-Markovian dynamics including non-exponential decay and atom-photon bound states exists in the system which results from the special (threshold-like) density of states (see Eq. (8)). When the atomic transition frequency lies in the bandgap ($\Delta_{c}<0$), the system exhibits photon-atom bound states and decaying states in the allowed band ($\Delta_{c}>0$). The dynamical behavior of the anisotropic system is almost the same as that of the isotropic system \cite{SCCheng09} except for the smaller probability of bound states and faster decaying behavior of decaying states. The dynamical difference of spontaneous emission in the two systems results from the different DOS in the two systems and the existence of diffusion field in the anisotropic system. As the atomic transition frequency moves from the bandgap to the allowed band, the density of states "seen" by the emitted photon is singularly large near band edge in the isotropic system and small in the anisotropic case. The singularity of density of states in isotropic system leads to the appearance of coherent propagating field while not large enough density of states results in the coexistence of incoherent diffusion field and coherent propagating field in anisotropic system. The energy transfer from the localized field to the diffusion field for the bound states of the anisotropic system leads to the smaller probability in excited level and coexisting energy of diffusion field and propagating field for decaying states results in faster decaying of the excited population. 

This dynamical difference in the isotropic and anisotropic systems is more obvious as the atomic transition frequency lying close to the edge of the PBG reservoir which is shown in Fig. 3.  As the atomic transition frequency lying close to the band edge, states in the isotropic system exhibit bound ($\Delta_{c}<0$) or slow decaying ($\Delta_{c}>0$) behavior. In the anisotropic system, however, almost all of these states display fast decaying behavior. That is, the bound states close to the band edge of the isotropic system change to the decaying states which lead to the appearance of SE in the anisotropic system. This result is consistent with the experimental observation in $Phys.$ $Rev.$ $Lett.$ \textbf{96}, 243902 (2006). In order to investigate the local optical properties of PCs, Barth et al. doped the PCs of artificial colloidal opals with CdSe/ZnS core-shell quantum dots whose emission frequency lies inside the forbidden gap and the linewidth was narrower than the width of the band gap. They demonstrated that the characteristic patterns of fluorescence image from different quantum dots carried information on the modification of the optical mode density which arose from the direction-dependent photonic stop band. The anisotropic band structure of the artificial colloidal opals, which corresponds to the direction-dependent photonic stop band, brought an extra angular anisotropy of fluorescence image that was detected by defocused wide-field image of the single CdSe/ZnS quantum dots. These quantum dots did not emit light if they were embedded inside a PC with weak anisotropy of band structure. That is, spontaneous emission appears only in the anisotropic photonic crystal as the emission frequency of the embedded quantum dots lies in the forbidden gap, but they did not emit light with bound states in the weak anisotropic PC system. The correctness of our results is validated by the presence of spontaneous emission in the anisotropic photonic crystal which differs from the prediction of the previous studies \cite{SYZhu00, YYang03A, YYang03}. Dynamical difference of the isotropic and anisotropic PC systems leads to the appearance of fluorescence image in the anisotropic system. Close to the band edge, DOS in the anisotropic case is nearly zero so that the incoherent diffusion field emerges to release some of the radiation energy. This diffusion field increases as the atomic frequency shifts from the bandgap to the band edge and decreases as the frequency shifts to the allowed band. The transferred radiation energy from localized field and propagating field reaches maximum at the band edge. The phenomenon of the dynamical behavior of the two systems arriving at the great difference close to the band edge illustrates the existence of the diffusion field in the anisotropic system.
 
In Fig. 4, we show how the curvature of the dispersion relation affects the dynamical behavior of the anisotropic system which has so far not been explored. The solid lines are plotted for the system with the larger curvature ($f=1$) of dispersion relation and dotted lines for the system with the smaller curvature ($f=0.8$). For the bound states ($\Delta_{c}<0$), the excited-state probability $P(t)$ of the system with the smaller curvature has the smaller value than that of the system with the larger curvature. On the other hand, the excited-state probability of the system with smaller curvature decays faster than that of the larger-curvature system for the decaying states ($\Delta_{c}>0$).  Without changing the units of energy (coupling constant $\beta$) and time ($1/\beta$), the dispersion relation of the smaller curvature has the larger DOS and coupling strength. As the atomic transition frequency moves from the bandgap to the allowed band, this larger DOS leads to the earlier appearance of the diffusion field that corresponds to the earlier energy transfer from the localized field to the diffusion field for the bound states and earlier coexisting energy of the diffusion field and the propagating field for the decaying states. This earlier energy transfer and coexistence resulting in the smaller value of probability in the bound states and fast decaying in the decaying states cause different optical behavior of the anisotropic system along different wavevectors.

\section{Conclusion}
We have used fractional calculus to solve the non-Markovian dynamics of the optical system consisting of a two-level atom coupled to a PBG reservoir with anisotropic one-band model. It is the first time in the anisotropic system that the analytical solution of the kinetic equation is obtained in terms of fractional exponential functions. The dynamical spontaneous emission has almost the same behavior as that of the isotropic system except for the smaller probability of the photon-atom bound states ($\Delta_{c}<0$) and the faster decaying rates for the decaying states ($\Delta_{c}>0$). The dynamical difference originates from the different DOS in the two systems and the existence of diffusion field in the anisotropic system. Not large enough DOS near band edge in the anisotropic system leads to the appearance of the incoherent diffusion field and energy transfer from the localized field to the diffusion field. The dynamical difference between the anisotropic and isotropic systems manifests itself more clearly as the atomic transition frequency lies close to the edge of the PBG reservoir. With the same atom-field coupling strength and detuning frequency in the forbidden gap, the bound states in the isotropic system turn into the unbound states in the anisotropic system. This change leads to the presence of SE in the anisotropic system that agrees with the experimental observation in Ref. \onlinecite{MBarth06} where spontaneous emission happens only in the strong anisotropic PCs but not in the weak anisotropic system for the emission frequency lying in the forbidden gap. The presence of spontaneous emission in the anisotropic photonic crystal validates the correctness of our results while illustrates the inconsistency with the prediction of the previous studies \cite{SYZhu00, YYang03A, YYang03}. The existence of the diffusion field in the anisotropic system is elucidated by the dynamical behavior of the two systems arriving at the great difference close to the band edge. We also investigated the new topic of how the curvature of the anisotropic dispersion relation affects the dynamical behavior of the anisotropic system. Without changing the units of energy and time, the dispersion relation of the smaller curvature has larger DOS and coupling strength which leads to the earlier appearance of the diffusion field and energy transfer. This earlier energy transfer resulting in the smaller and faster-decaying probability causes the different optical behavior along different wavevectors in the anisotropic PC system.

\begin{acknowledgments}
We would like to gratefully acknowledge partially financial support from the National Science Council (NSC), Taiwan under Contract Nos. NSC-97-2811-E-009-023, NSC-96-2628-M-009-001-MY3, NSC-96-2112-M-034-002-MY3, and NSC-96-2628-M-009-001-MY3.
\end{acknowledgments}
\bibliography{TwoLevelAtomInAPC}

\begin{figure}
\includegraphics[bb=101 540 524 729, width=15.0cm, totalheight=7.5cm, clip]{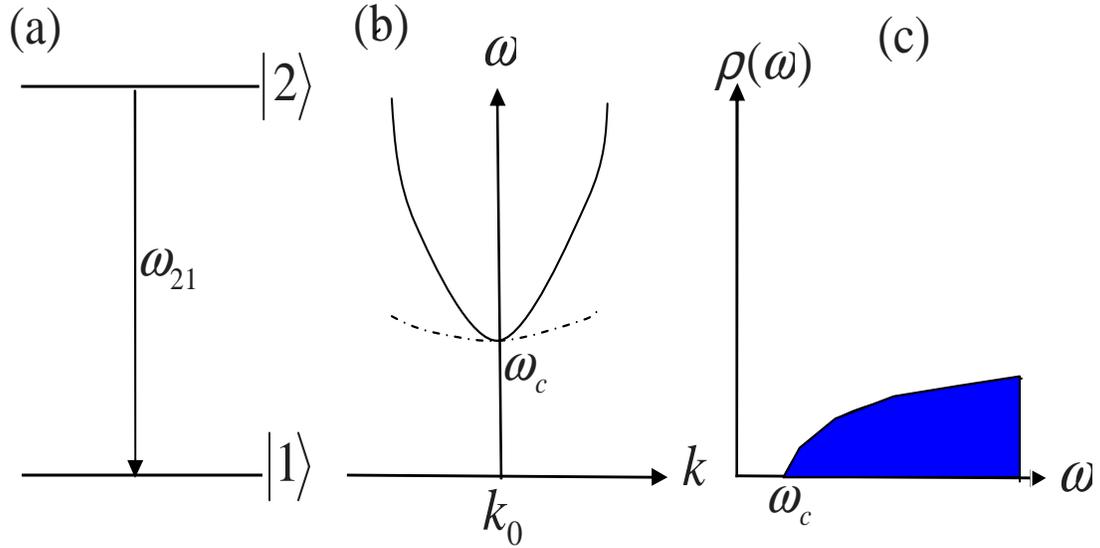}\label{Fig1}
\caption{(a) A two-level atom with excited state $\left|2\right\rangle$ and ground state $\left|1\right\rangle$. The transition frequency $\omega_{21}$ is nearly resonant with the frequency range of the PBG reservoir. (b) Directional dependent dispersion relation near band edge with edge frequency $\omega_{c}$ . (c) DOS of the anisotropic one-band effective mass model. }
\end{figure}
%%%%%%%%%%%%%%%%%%%%%%%%%%%%%%%%%%%%%%%%%%%%%%%%%%%%%%%%%%%%%%%%%%%%%%%%%%%%%%%%%%

\begin{figure}
\includegraphics[bb=18 19 303 232, width=12.0cm, totalheight=9.5cm, clip]{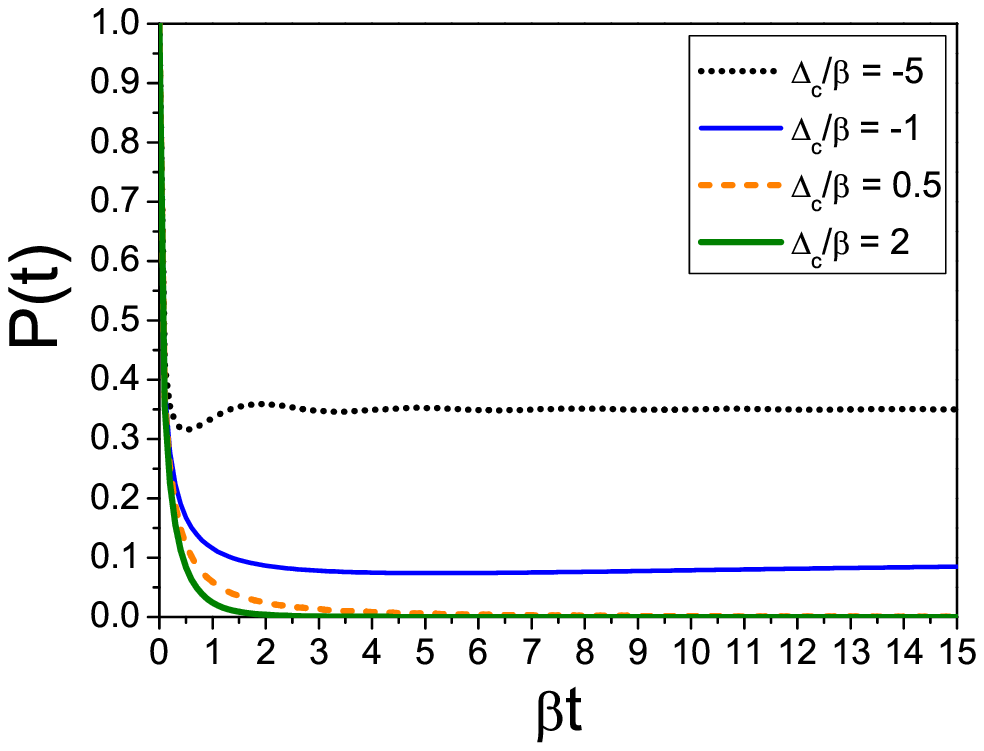} \label{Fig2}
\caption{ Dynamics of SE of the anisotropic PC system as a function of $\beta t$ for various values of the atomic detuning frequency ($\Delta_{c}=\omega_{21}-\omega_{c}$). States of $\Delta_{c}/\beta<0$ inside the band gap and states of $\Delta_{c}/\beta>0$ within the allowed band.}
\end{figure}

%%%%%%%%%%%%%%%%%%%%%%%%%%%%%%%%%%%%%%%%%%%%%%%%%%%%%%%%%%%%%%%%%%%%%%%%%%%%%%%%%%
\begin{figure}
\includegraphics[bb=18 19 303 232, width=12.0cm, totalheight=9.5cm, clip]{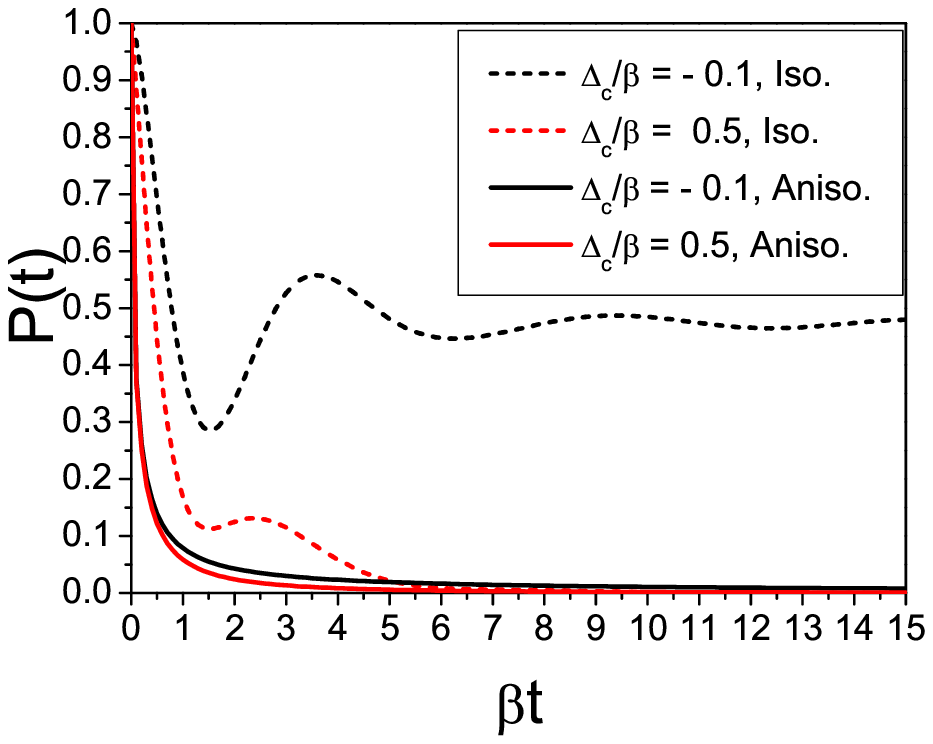}\label{Fig3}
\caption{Dynamics of SE of the anisotropic (Aniso., solid lines) and isotropic (Iso., dashed lines) systems with detuning frequency close to the band edge of PBG reservoir ($\Delta_{c}=\omega_{21}-\omega_{c}\rightarrow 0$).}
\end{figure}

%%%%%%%%%%%%%%%%%%%%%%%%%%%%%%%%%%%%%%%%%%%%%%%%%%%%%%%%%%%%%%%%%%%%%%%%%%%%%%%%%%
\begin{figure}
\includegraphics[bb=16 15 302 238, width=12.0cm, totalheight=9.5cm, clip]{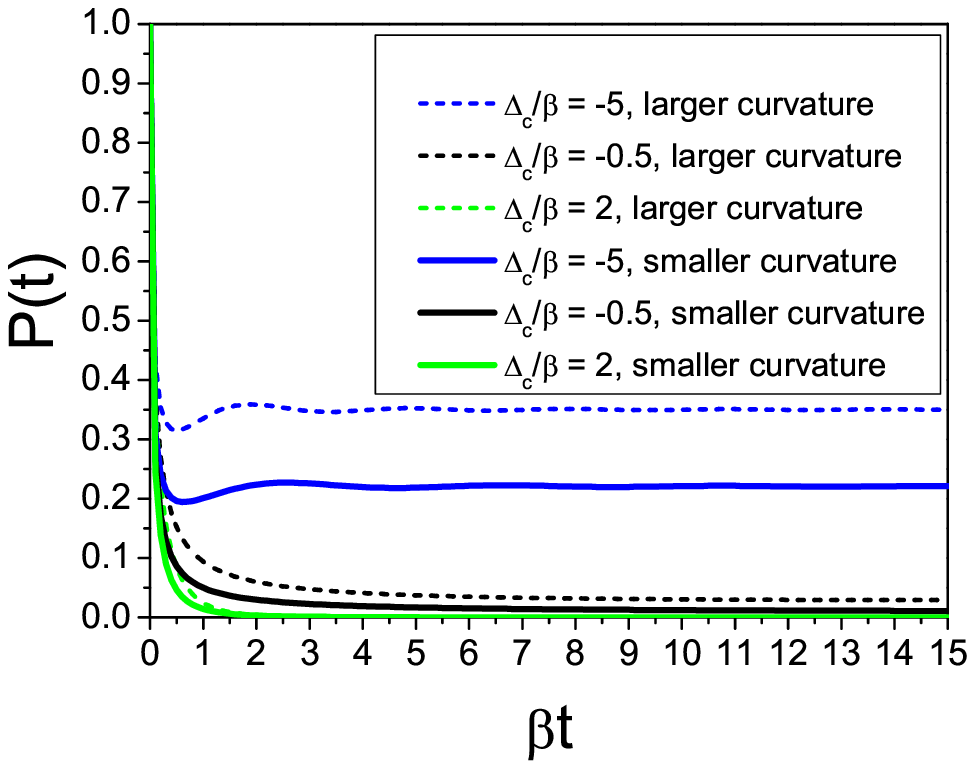}\label{Fig4}
\caption{Dynamics of SE of the anisotropic system with two kinds of curvatures in the dispersion relation. Solid lines for the system with smaller-curvature dispersion relation and dashed lines for the larger-curvature system. }
\end{figure}

\end{document}